\documentclass[twocolumn,english,floatfix, aps,arxiv,superscriptaddress]{revtex4-1}
\usepackage{grffile}
\usepackage{amsmath}
\usepackage[latin9]{inputenc}
\usepackage{color}
\usepackage{babel}
\usepackage{graphicx}
\usepackage{cancel}
\PassOptionsToPackage{normalem}{ulem}
\usepackage{ulem}
\usepackage[unicode=true, bookmarks=false, breaklinks=false,pdfborder={0 0 1},colorlinks=false] {hyperref}

\begin{document}

\preprint{APS/123-QED}

\title{Revealing the competition between charge-density wave and superconductivity in CsV$_3$Sb$_5$ through uniaxial strain}

\author{Tiema Qian}
\affiliation{Department of Physics and Astronomy and California NanoSystems Institute, University of California, Los Angeles, Los Angeles, CA 90095, USA}

\author{Morten H. Christensen}
\affiliation{Niels Bohr Institute, University of Copenhagen, 2100 Copenhagen, Denmark}

\author{Chaowei Hu}
\affiliation{Department of Physics and Astronomy and California NanoSystems Institute, University of California, Los Angeles, Los Angeles, CA 90095, USA}

\author{Amartyajyoti Saha}
\affiliation{School of Physics and Astronomy, University of Minnesota, Minneapolis, MN 55455, USA}
\affiliation{Department of Chemical Engineering and Materials Science, University of Minnesota, MN 55455, USA}

\author{Brian M. Andersen}
\affiliation{Niels Bohr Institute, University of Copenhagen, 2100 Copenhagen, Denmark}

\author{Rafael M. Fernandes}
\affiliation{School of Physics and Astronomy, University of Minnesota, Minneapolis, MN 55455, USA}

\author{Turan Birol}
\affiliation{Department of Chemical Engineering and Materials Science, University of Minnesota, MN 55455, USA}

\author{Ni Ni}
\email{Corresponding author: nini@physics.ucla.edu}
\affiliation{Department of Physics and Astronomy and California NanoSystems Institute, University of California, Los Angeles, Los Angeles, CA 90095, USA}

\date{\today}

\begin{abstract} 

In this paper we report the effect of uniaxial strain $\varepsilon$ applied along the crystalline $a$ axis on the newly discovered kagome superconductor CsV$_3$Sb$_5$. At ambient conditions, CsV$_3$Sb$_5$ shows a charge-density wave (CDW) transition at $T_{\rm CDW}=94.5$ K and superconducts below $T_C = 3.34$ K. In our study, when the uniaxial strain $\varepsilon$ is varied from $-0.90\%$ to $0.90\%$, $T_C$ monotonically increases by $\sim 33\%$ from 3.0 K to 4.0 K, giving rise to the empirical relation $T_C (\varepsilon)=3.4+0.56\varepsilon+0.12\varepsilon^2$. On the other hand, for $\varepsilon$ changing from $-0.76\%$ to $1.26\%$, $T_{\rm CDW}$ decreases monotonically by $\sim 10\%$ from 97.5 K to 87.5 K with $T_{\rm CDW}(\varepsilon)=94.5-4.72\varepsilon-0.60\varepsilon^2$. The opposite response of $T_C$ and $T_{\rm CDW}$ to the uniaxial strain suggests strong competition between these two orders. Comparison with hydrostatic pressure measurements indicate that it is the change in the $c$-axis that is responsible for these behaviors of the CDW and superconducting transitions, and that the explicit breaking of the sixfold rotational symmetry by strain has a negligible effect. Combined with our first-principles calculations and phenomenological analysis, we conclude that the enhancement in $T_C$ with decreasing $c$ is caused primarily by the suppression of $T_{\rm CDW}$, rather than strain-induced modifications in the bare superconducting parameters. We propose that the sensitivity of $T_{\rm CDW}$ with respect to the changes in the $c$-axis arises from the impact of the latter on the trilinear coupling between the $M_1^+$ and $L_2^-$ phonon modes associated with the CDW. Overall, our work reveals that the $c$-axis lattice parameter, which can be controlled by both pressure and uniaxial strain, is a powerful tuning knob for the phase diagram of CsV$_3$Sb$_5$.

\end{abstract}

\maketitle

\section{Introduction}

The interplay between superconductivity (SC) and charge-density waves (CDW) has a long history spanning several different classes of materials \cite{GabovichCDW&SC,zhu2015classification}. In metals, while SC is a Fermi surface instability, CDW can arise due to nesting of the Fermi surface, lattice instabilities, or the electron-phonon interaction. As a result, the nature of the coupling between SC and CDW, from competing to cooperative, can be quite rich.
For instance, in Cu$_x$TiSe$_2$ or pressurized TiSe$_2$, the superconducting transition temperature ($T_C$) reaches its maximum when the CDW state is suppressed completely \cite{TiSe2competing, CuxTiSe2competing}, suggesting a possible link between CDW fluctuations and the formation of Cooper pairs. In cuprates, CDW fluctuations are observed to be suppressed below the onset of SC, indicative of competition between the two phases \cite{Wu2011,Chang2012}. On the other hand, in materials such as pressurized 1T-TaS$_2$ and 2H-NbSe$_2$ \cite{Independent1T-TaS2,IndependentNbSe2}, SC seems to be little affected by the suppression of CDW. 

\begin{figure}
    \centering
    \includegraphics[width=\columnwidth]{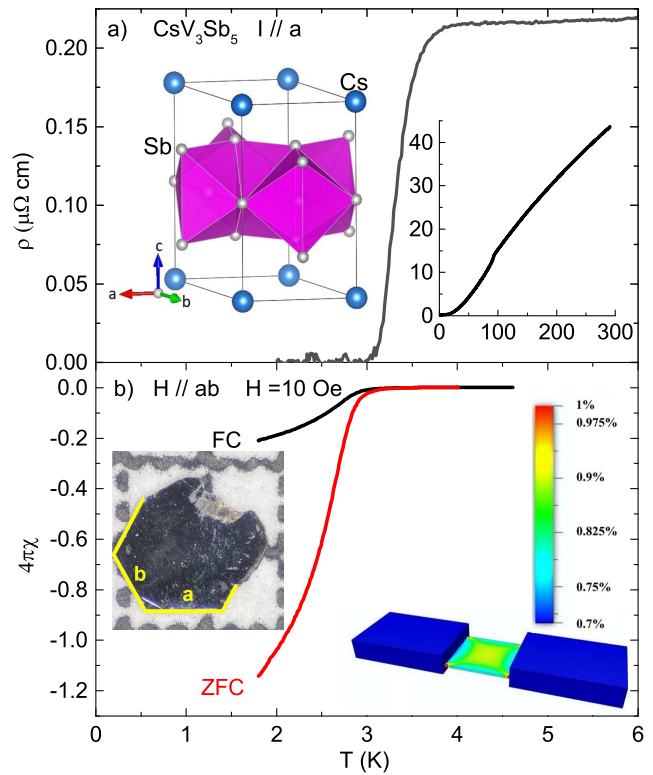}
    \caption{a) The zoomed-in temperature-dependent electrical resistivity $\rho_{xx}(T)$ near the superconducting transition with the current along the $a$ axis. Left inset: The crystal structure of CsV$_3$Sb$_5$. Right inset: $\rho(T)$ from 2 K to 300 K. b) The ZFC and FC susceptibilities measured at 10 Oe with $H || ab$. Left inset: a piece of single crystal against a 1-mm scale with the $a$ and $b$ axes labeled. Right inset: the mapping of the actual strain on the sample, see text for more details.}
    \label{cray}
\end{figure}

Recently, a family of quasi-two-dimensional kagome materials, $A$V$_3$Sb$_5$ ($A =$ K, Rb, and Cs), has been discovered \cite{KagomeDiscovery}, sparking the interest of the community due to the presence of SC, CDW and non-trivial band structure
\cite{KVSbSuperconduct, ScVSbSc/CDW, RbV3Sb5SC, chen2021roton, ScVSbSc/CDW,liang2021three, tan2021charge,li2021observation}. As shown in the inset of Fig.~\ref{cray}(a), at room temperature, CsV$_3$Sb$_5$ crystallizes in the hexagonal space group $P 6/m m m$ with alternating Cs layers and V$_3$Sb$_5$ layers made of face-sharing VSb$_6$ octahedra. Of particular importance, the V atoms form a kagome lattice, which has been proposed to be the major structural ingredient responsible for the emergent phenomena of CDW, with transition temperature $T_{\rm CDW}\sim94$ K, and of SC, with $T_C\sim3$ K. The structure of the CDW phase remains widely debated, with studies reporting unidirectional CDW \cite{zhao2021cascade,liang2021three,chen2021roton}, a three-dimensional CDW with a  2$\times$2$\times$2 superstructure  \cite{ortiz2021fermi} or a 2$\times$2$\times$4 superstructure \cite{liang2021three}, a chiral CDW ~\cite{Jiang2021,mielke2021time,shumiya2021tunable}, and a CDW that breaks the sixfold rotational symmetry of the kagome lattice~\cite{zhao2021cascade,li2021rotation}.

A rich interplay between CDW and SC was observed in CsV$_3$Sb$_5$ under external hydrostatic pressure~\cite{zhao2021nodal,chen2021double,yu2021unusual,zhang2021pressure,chen2021highly,tsirlin2021anisotropic}. Upon increasing the pressure up to 10 GPa, $T_C$ and $T_{\rm CDW}$ were found to compete with each other, leading to a SC dome in the temperature-pressure phase diagram, with the maximum $T_C\sim 8$ K occurring at a pressure of 2 GPa, where the CDW order is completely suppressed. Furthermore, a dip in the SC dome was observed at $\sim 1$ GPa, concurrent with a possible commensurate to nearly-commensurate CDW transition~\cite{yu2021unusual}. When the pressure was further increased, an additional SC dome with a maximum $T_C$ of 5 K appears and persists up to 100 GPa, the maximum pressure measured. Despite the rapidly-evolving understanding of the CDW and the competition between CDW and SC, the nature of the SC state remains unsettled. While thermal conductivity measurements suggest nodal SC \cite{zhao2021nodal}, penetration depth measurements indicate nodeless SC \cite{duan2021nodeless}. 

To better understand the interplay between SC and CDW, we investigate their responses to uniaxial strain applied along the $a$ axis in CsV$_3$Sb$_5$. Comparing to hydrostatic pressure, which equally compresses the lattice along all directions, uniaxial strain not only explicitly breaks the sixfold rotational symmetry of the lattice, but it can both compress and stretch the lattice along a certain direction. It has been employed previously as a powerful tool to tune and detect exotic phases in both topological~\cite{mutch2019evidence,stern2017surface} and strongly correlated systems~\cite{chu2012divergent,hicks2014strong,kissikov2018uniaxial}. The linear-dominated monotonic dependence of $T_{\rm CDW}$ with strain reveals that the symmetry-breaking effect on the CDW is negligible. Instead, the $T_{\rm CDW}$ and $T_c$ data for tensile $a$-axis strain quantitatively agree with the hydrostatic pressure data when both are plotted as a function of the $c$-axis compression. This strongly suggests that the structural parameter to which CsV$_3$Sb$_5$ is most sensitive is the $c$-axis lattice parameter. Moreover, the ratio of change of $T_C$ with respect to $T_{\rm CDW}$ is almost identical to that seen in the pressure experiments. 

Combined with first-principles calculations and a phenomenological analysis, we conclude that the enhancement of $T_C$ with the tensile $a$-axis strain is likely entirely due to the suppression of the competing CDW order, rather than an independent change in the bare superconducting parameters, like the density of states. Such a strong competition between CDW and SC is indicative of phases competing for similar electronic states. We further propose that the suppression of $T_{\rm CDW}$ with the tensile $a$-axis strain is associated with a $c$-axis induced change in the trilinear coupling between the CDW order parameters with wave-vectors at the $M$ and $L$ points of the Brillouin zone.

This paper is organized as follows. In Sec. II, we describe our experimental and theoretical methods, including how we determine the magnitude of the applied strain. In Sec. III, we present the electronic resistivity measurements under ambient and strained conditions. A comparison between the current work and the pressure measurements in literature is made. In Sec. IV, first-principles calculations and a phenomenological analysis are employed to interpret the experimental data. We conclude our paper in Sec. V. 

\section{Methods}

CsV$_3$Sb$_5$ was synthesized with the Cs-Sb flux method~\cite{KagomeDiscovery}. Cs, V, and Sb elements were loaded into an alumina crucible at the molar ratio of 20:16.7:63.3 and subsequently sealed in a quartz ampule under 1/3 atm of argon. The quartz was heated to 1000 $^{\circ}$C in 10 hours, where it dwelled for 20 hours, and then cooled to 800 $^{\circ}$C in 20 hours, followed by a further cooling to 600 $^{\circ}$C in one week. Finally, the furnace was turned off at 600 $^{\circ}$C and the tube was taken out at room temperature. Millimeter-sized plate-like crystals can be separated once the product is immersed in water for hours in the fume hood to remove the flux.

The phase and purity of the single crystals were confirmed by the X-ray diffractions via a PANalytical Empyrean diffractometer (Cu K$\alpha$). The electrical resistivity and magnetic properties were measured in a Quantum Design DynaCool Physical Properties Measurement System and Magnetic Properties Measurement System, respectively. Uniaxial strain was applied along the $a$-axis using a home-built three-piezostack strain apparatus \cite{hicks}. A single crystal was carefully cut into a rectangular resistivity bar with the $a$-axis as the long side. This bar was then glued across the gap of the strain apparatus using Stycast. To minimize the strain gradients between the top and the bottom surfaces, extra care was made so that both ends of the bar were completely submerged in Stycast. A foil strain gauge glued on one of the piezostacks was used to determine the value of strain, $\varepsilon_{\rm piezo}$. Then the total strain induced by the apparatus was estimated as $\varepsilon_{\rm total} = 2 \times \frac{L}{l} \times \varepsilon_{\rm piezo}$, where $L$ is the length of the piezo stack (9 mm) and $l$ is the width of the apparatus gap (0.25 mm). Finally, $\varepsilon$, the actual average strain induced on the samples can be written as $A\varepsilon_{\rm total}$ where $A$ is a constant and determined via the Finite Element Analysis (FEA) using Autodesk Fusion 360. We modelled a 1 mm $\times$ 0.23 mm $\times$ 0.01 mm crystal glued by Stycast across our apparatus with a gap size of 0.25 mm. The Stycast glue is modeled as 0.02 mm between the sample and the strain apparatus and 0.05 mm above the sample. $A$ was calculated to be ~0.9 on the portion of sample that is measured, as shown in the right inset of Fig. 1(b).  
 
\begin{figure}
    \centering
    \includegraphics[width=3.3in]{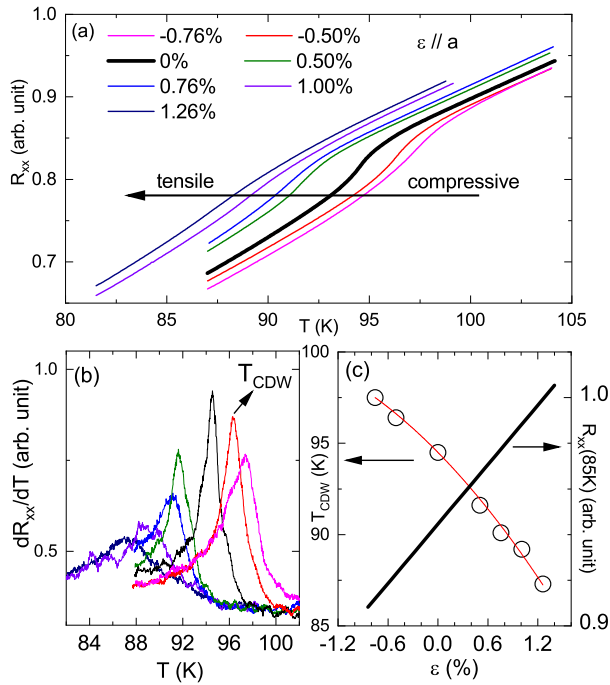}
    \caption{Strain dependence of $T_{\rm CDW}$ of a single crystal of CsV$_3$Sb$_5$. (a) Temperature dependence of the in-plane resistance of CsV$_3$Sb$_5$ near the CDW transition measured at different strain values. Negative values denote compressive strain, whereas positive values, tensile strain. (b) The temperature-derivative of the resistance, $dR_{xx}/dT$, whose peaks determine $T_{\rm CDW}$. (c) $T_{\rm CDW}$ and resistance as a function of strain applied along the $a$-axis. Resistance is linearly proportional to the strain applied.}
    \label{tcdw}
\end{figure}

To calculate the density of states and phonon frequencies, first principles density functional theory calculations were performed using the Vienna Ab Initio Software Package (VASP) and the projector augmented waves formalism \cite{Kresse1993,Kresse1996CMS,Kresse1996PRB}. The exchange-correlation energy was approximated using the PBEsol functional and without a +U correction \cite{Perdew2008}. Reciprocal space k-grids with a density of a point per $\sim0.012\times 2\pi$\AA$^{-1}$ were used in all calculations. The plane wave cutoff of 350~eV, and Cs and V potentials with $s$ semi-core states treated as valence states were employed. Phonon calculations were performed using the frozen phonons technique. Since Fermi surface smearing is found to have an effect on the phonon frequencies, a Gaussian smearing with 1~meV width was used. The smearing does not make a qualitative difference in the densities of states, however, the phonon frequencies are dependent on the smearing as discussed in Ref. \cite{christensen2021theory}.

Uniaxial strain along the $a$-direction was simulated by fixing the magnitude of the lattice parameter $a$, and relaxing both the ionic positions and the other two lattice parameters. For results presented in Fig. \ref{fig:dft}, which shows the trends of the phonons under decreasing $c$, only the internal coordinates of the atoms were relaxed.

\section{Experimental results}
Samples grew in hexagonal plates with clear as-grown edges, as shown in the left inset of Fig. 1(b). The X-ray pattern taken on the surface of the plate can be indexed by the (00$l$) reflections, indicating that the as-grown edges marked by the yellow lines are the crystalline $a$ and $b$ axes. 

The temperature-dependent resistivity $\rho_{xx}(T)$ at ambient conditions is shown in the right inset of Fig. 1(a). Following a resistivity drop at 94.5 K, the sample enters the superconducting state below 3.4 K. These features are consistent with the values provided in the literature for the temperatures where the CDW and SC transitions are observed. The SC transition can be
seen more clearly in the low-field susceptibility data presented in Fig. 1(b). The black curve is the field-cooled (FC) data while the red curve is the zero-field-cooled (ZFC) data. The large diamagnetic signal can be seen in both the ZFC and FC case, suggesting a large shielding fraction implying bulk superconductivity.

The right inset of Fig.~1(b) shows the map of the actual strain applied on the sample via FEA. Based on it, the average strain applied in the measured sample portion is $\varepsilon=0.9\varepsilon_{\rm total}$. In our experiment, before the resistivity bar cracked due to the applied strain, the setup successfully applied uniaxial strain from $-0.90\%$ (i.e. compressive strain) to $0.90\%$ (i.e. tensile strain) around 2~K and from $-0.76\%$ to $1.26\%$ around 85~K.

\begin{figure}
    \centering
    \includegraphics[width=3.4in]{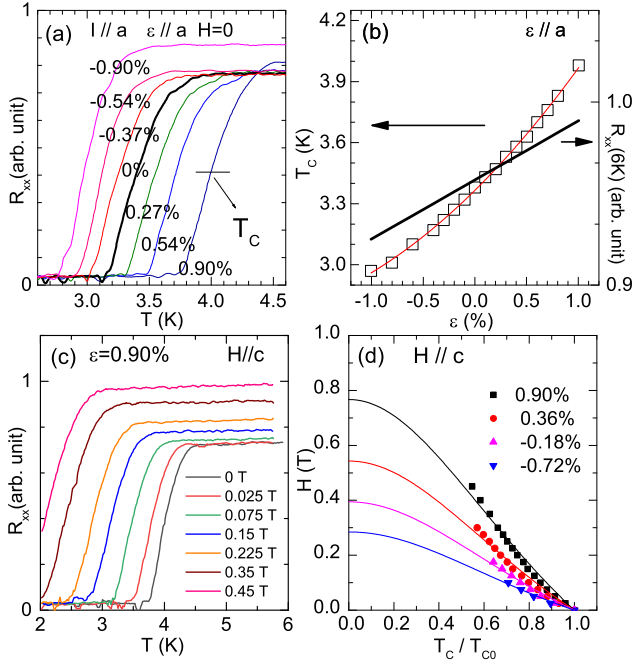}
    \caption{Strain dependence of $T_C$ of CsV$_3$Sb$_5$. (a) Temperature dependence of the in-plane resistance $R_{xx}$ of CsV$_3$Sb$_5$ near the superconducting transition. (b) Strain dependence of the resistance at $6$ K and of $T_C$, which is determined by the 50\%-resistance criterion. Resistance is linearly proportional to the strain applied. (c) Magnetic field dependence of the in-plane resistance in the presence of an applied 0.90\% tensile strain. (d) $H_{c2}$ diagram of CsV$_3$Sb$_5$ subjected to different strain values. Solid lines are fittings to the Ginzburg-Landau model.}
    \label{tc}
\end{figure}

 \begin{figure*}
    \centering
    \includegraphics[width=7in]{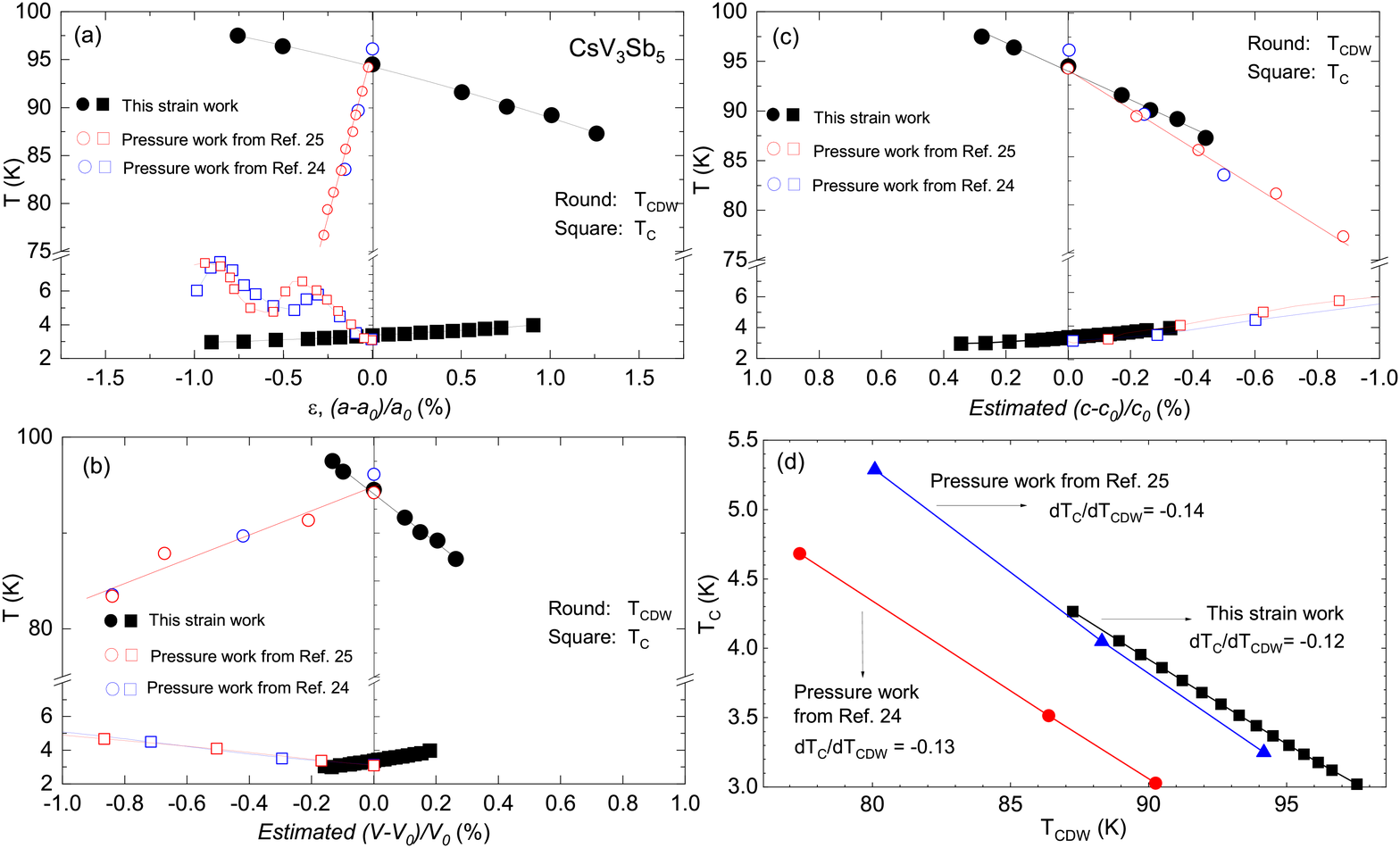}
    \caption{Comparison between the phase diagrams obtained from uniaxial strain measurements (this work) and previous hydrostatic pressure measurements \cite{yu2021unusual,chen2021double} (a) The effect of the change in lattice parameter $a$ on $T_{\rm CDW}$ and $T_C$. The relationship between the lattice parameters and pressure was obtained from Ref. \cite{zhang2021pressure}. (b) The effect of the change in volume $V$ on $T_{\rm CDW}$ and $T_c$. The response of the lattice parameter $c$ with respect to the strain was extracted from DFT calculations. (c) The effect of the change in $c$ on $T_{\rm CDW}$ and $T_c$. (d) $T_C$ plotted as a function $T_{\rm CDW}$ for both pressure and uniaxial strain experiments.}
    \label{tcdw}
\end{figure*}

Figure 2(a) shows the resistance data $R_{xx}(T)$ around the CDW transition under various uniaxial strain values. As one can see, the resistance drop associated with the CDW transition moves to lower temperatures upon the application of tensile strain whereas it goes to higher temperatures when the sample is compressed. To determine $T_{\rm CDW}$ under strain, we use the peak position of the derivative $d R_{xx}/dT$, as shown in Fig. 2(b). Figure 2(c) shows the results for the strain-dependent $T_{\rm CDW}(\varepsilon)$. It monotonically decreases when the sample expands along the $a$ axis, being 97.5 K when the $a$-axis is strained by $-0.76\%$, 94.5 K for no applied strain, and 87.3 K when the $a$-axis expands by $1.26\%$. A polynomial fitting to $T_{\rm CDW}(\varepsilon)$ gives $T_{\rm CDW}(\varepsilon)=94.5-4.72\varepsilon-0.60\varepsilon^2$, revealing a dominant linear term.

Figure 3(a) shows the $R_{xx}(T)$ data around the superconducting transition under applied strain. $T_C$ increases with compressive strain and decreases for tensile strain. $T_C$ is determined using the $50\%$-resistance criterion, as shown in Fig. 3(a). Figure 3(b) summarizes the strain-dependent $T_C(\varepsilon)$. It is 3.0 K at $-0.90\%$ strain, 3.4 K for no applied strain, and 4.0 K at $0.90\%$ strain. A polynomial fitting to $T_C(\varepsilon)$ gives $T_C (\varepsilon)=3.4+0.56\varepsilon+0.12\varepsilon^2$.

Figure 3(c) shows a representative example of the  behavior of $R_{xx}(T)$ in a strained sample ($\varepsilon=0.90\%$ in this case) under different magnetic fields applied along the $c$-axis, $H\parallel c$. As expected, the SC transition is suppressed with increasing fields. The field dependence of $T_C$ under different strains is summarized in Fig. 3(d). The zero-temperature upper critical field $H_{c2}^{\parallel c}(0)$ can then be estimated by fitting the $H_{c2}(T)$ data with the empirical Ginzburg-Landau equation, $H_{c2}(T) = H_{c2}(0)(1-t^2)/(1+t^2)$, where $H_{c2}(0)$ is the zero-temperature upper critical field and $t=T/T_C$ is the reduced temperature. The fitting curves are shown in Fig. 3(d) as solid lines. We see that $H_{c2}^{\parallel c}(0)$ has a strong strain dependence, increasing to 0.77 T when $0.90\%$ strain is applied and decreasing to 0.29~T when $-0.72\%$ strain is applied.

Although strain was applied along the $a$-axis, the sample will deform (unequally) in all three directions according to the Poisson ratios. To obtain a better understanding of which of the lattice parameters controls $T_{\rm CDW}$ and $T_C$ in our experiments, DFT calculations were performed to calculate the changes in the lattice parameter $c$ and in the volume $V$ under our experimental conditions. The data was then compared to previous hydrostatic pressure measurements, in which case all three directions are compressed equally. Using the lattice parameter changes with respect to pressure reported in Ref.~\onlinecite{zhang2021pressure}, the relationship between the CDW/SC transition temperatures and the changes in different lattice parameters can be compared for our uniaxial strain experiment and the hydrostatic pressure experiments of Refs.~\onlinecite{chen2021double,yu2021unusual}. 

In Figs. 4(a), (b) and (c), we plot the transition temperatures $T_{\rm CDW}$ and $T_C$ as a function of the change in the lattice parameter $a$, volume $V$ and lattice parameter $c$, respectively, for our work and for the pressure data of Refs.~\onlinecite{chen2021double,yu2021unusual}. Figure 4(a) shows that, in the case of hydrostatic pressure, a decreasing $a$ leads to a decrease in $T_{\rm CDW}$ and an initial increase in $T_{C}$. Meanwhile, in the case of uniaxial strain, upon decreasing $a$, $T_{\rm CDW}$ increases and $T_{C}$ decreases. The opposite response of $T_{\rm CDW}$ and $T_C$ with respect to changes in $a$ via two different experimental techniques suggests that $a$ is not the primary lattice parameter tuning the CDW and SC. A similar conclusion can be drawn from Fig. 4(b), indicating that the volume $V$ is not the main tuning parameter either. On the other hand, as shown in Fig. 4(c), for both the pressure and the uniaxial strain data, $T_{\rm CDW}$ decreases and $T_C$ increases with decreasing $c$. Clearly, the agreement is not only qualitative, but also quantitative: the ratio of change of $T_{\rm CDW}$ ($T_C$) with respect to the $c$-axis compression is nearly the same for both uniaxial strain and pressure data. These observations provide strong evidence that $c$ is the primary lattice parameter responsible for tuning the CDW and SC transitions.

In both pressure and strain measurements, in the region of $c$-axis compression, $T_{\rm CDW}$ is suppressed while $T_C$ is enhanced. In the strain measurements, which can also assess the region of $c$-axis expansion, we further notice that when $T_{\rm CDW}$ is enhanced, $T_C$ is suppressed. This competing relationship between CDW and SC indicates that both phenomena are affected by similar electronic states. Figure 4(d) presents a plot of $T_C$ as a function of $T_{\rm CDW}$ for both pressure and strain experiments. In all cases, $T_{C}$ depends linearly on $T_{\rm CDW}$ with a negative slope that varies weakly from $-0.12$ to $-0.14$, within the experimental error. Combined with Fig. 4(c), this result suggests the reason why $T_C$ increases upon the application of pressure or tensile $a$-axis strain is because the competing CDW order is suppressed due to the change in the $c$-axis lattice parameter.

Last but not least, one important difference between uniaxial strain and hydrostatic pressure is that the former breaks the (sixfold) rotational symmetry of the lattice, but the latter does not. The fact that $T_C$ and $T_{\rm CDW}$ change in the same way regardless of whether pressure or uniaxial strain is applied, as long as the $c$-axis compression is the same in both cases, indicates that the impact of the explicit breaking of the lattice symmetry is negligible compared to the effect arising from the change in the $c$-axis lattice parameter.

\section{Discussion}

\begin{figure}
    \centering
    \includegraphics[width=\columnwidth]{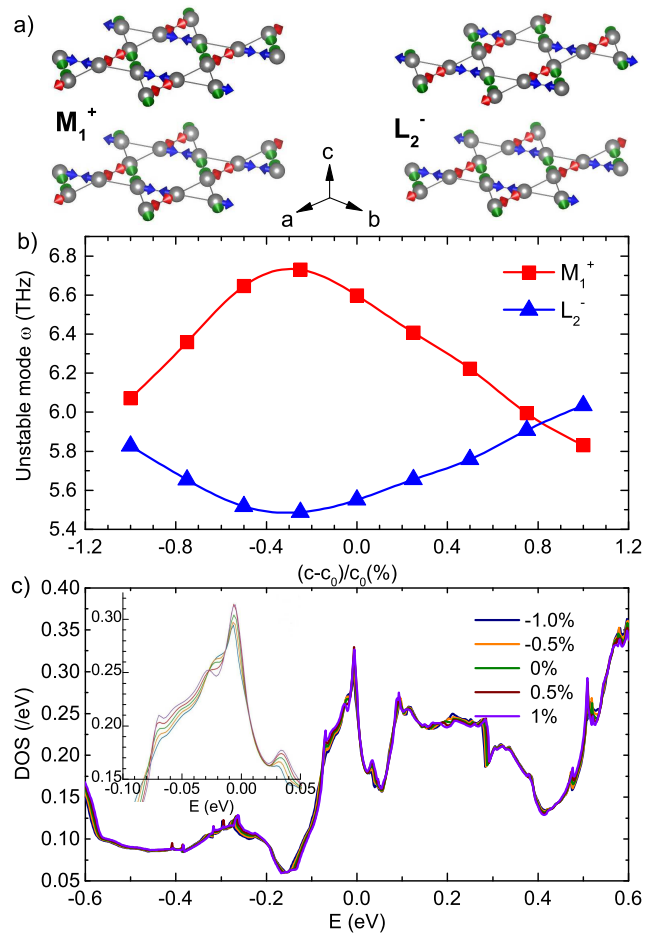}
    \caption{(a) Left: the displacement pattern of the $M_1^+$ phonon mode. For clarity, only the vanadium ions and two consecutive kagome layers are shown. Red, blue, and green arrows correspond to displacements according to the $M_1$, $M_2$ and $M_3$ components. Right: Same as the left panel, but for the $L_2^-$ mode. (b) Unstable phonon frequencies as a function of the $c$ lattice parameter from first principles calculations. Because the squared phonon frequencies are negative for unstable modes, we plot the square root of the absolute value of the squared frequencies. (c) Electronic density of states (DOS) of CsV$_3$Sb$_5$, calculated from DFT, in the high temperature, high symmetry phase, as a function of the lattice parameter $c$. The inset is a zoom of the low-energy part of the spectrum. The van Hove singularity peak below the Fermi level is suppressed under compression of the $c$-axis.}
    \label{fig:dft}
\end{figure}

We have shown that the $c$-axis lattice parameter is the dominant structural parameter that controls the changes of $T_{\rm CDW}$ and $T_C$ upon the application of uniaxial strain or pressure. To shed light on the microscopic mechanism of this effect, we calculate the strain-dependent phonon frequencies and density of states (DOS) using DFT in the non-CDW phase. Similar to Ref.~\onlinecite{ratcliff2021coherent}, our DFT results reveal CsV$_3$Sb$_5$ to have at least two unstable phonon modes at the $M$ and $L$ points of the Brillouin zone, which are associated with the CDW transition (see also Ref. \onlinecite{christensen2021theory}). These phonon modes transform as the $M_1^+$ and $L_2^-$ irreducible representations of the space group, as shown in Fig.~\ref{fig:dft}(a). 

In Fig.~\ref{fig:dft}(b), we display the behavior of these unstable phonon frequencies as a function of changes in the lattice parameter $c$. We find that the frequencies associated with the two modes show opposite and non-monotonic trends as a function of $c$ in the region $-1\%\leq (c-c_0)/c_0 \leq 1\%$ (see Fig. 4(c)). Similar calculations repeated with different electronic smearing parameters indicate that Fig.~\ref{fig:dft} sensitively depends on how the electronic structure is treated. Such an observed sensitivity supports the view that the electronic degrees of freedom are responsible for the unstable phonons observed in DFT~\cite{christensen2021theory}. While a precise quantitative prediction is obscured due to the dependence on the electronic smearing, the non-monotonic features and the opposite trends of the two modes remain robust. So qualitatively, if we associate the unstable phonon frequencies to the energy scale of the CDW transition, the opposite trends in the $L$ and $M$ modes, which will both condense in the $2\times2\times2$ CDW state, seems difficult to reconcile with the monotonic, nearly linear suppression of $T_{\rm CDW}$ with $(c-c_0)/c_0$ seen experimentally. 

To further elucidate this issue, we employ the Landau free-energy expansion for the CDW transition in $A$V$_3$Sb$_5$~\cite{christensen2021theory}. The two unstable phonon modes are three-fold degenerate because the hexagonal symmetry of the lattice gives rise to three distinct $M$ and $L$ wave-vectors at different faces of the Brillouin zone. We denote the CDW order parameters associated with different $M$ wave-vectors as $M_1$, $M_2$, and $M_3$, and similarly, we use $L_1$, $L_2$, and $L_3$ to denote the order parameters with $L$ wave-vectors. An illustration of these different order parameters can be found in Fig.~\ref{fig:dft}(a) in terms of distortions of the bonds connecting the V atoms. Since the out-of-plane component of the $M$ wave-vector is zero, the displacement of atoms on two consecutive layers are in phase. On the other hand, because of the out-of-plane component of the wave-vector $L$, the displacement of atoms on two consecutive layers are out of phase. To quadratic order, the free energies of these two order parameters are decoupled and given simply by:
\begin{equation}
\mathcal{F}^{(2)} = \alpha_M \left(M_1^2+M_2^2+M_3^2\right) + \alpha_L \left(L_1^2+L_2^2+L_3^2\right)
\end{equation}
The leading coupling between the two order parameters appears in cubic order as the trilinear coupling \cite{christensen2021theory}:
\begin{equation} \label{equ:landau}
\mathcal{F}^{(3)} = \gamma_{ML} \left(M_1 L_2 L_3 + M_2 L_3 L_1 + M_3 L_1 L_2 \right)
\end{equation}
This term is allowed since adding up one of the three $M$ wave-vectors with the two ``opposite" $L$ wave-vectors gives zero~\cite{christensen2021theory}. 


The Landau coefficients $\alpha_M$, $\alpha_L$, and $\gamma_{ML}$ are material-specific and, quite generically, expected to depend on the lattice parameter $c$. While they can in principle be calculated by DFT, this is an involved calculation that is beyond the scope of this work. Notwithstanding the aforementioned issues with the calculated DFT phonon frequencies, we can still gain some insight from the trends shown in Fig.~\ref{fig:dft}(b). Generally, the quadratic coefficient of a Landau free-energy expansion is positive in the disordered state and negative in the ordered state. Thus, in our case, we expect the quadratic coefficients $\alpha_M$ and $\alpha_L$ to be negative in the $2\times2\times2$ CDW state. Because the squared frequencies of the $M^{+}_1$ and $L_2^-$ phonon modes are positive in the non-CDW state and negative in the CDW state, it is reasonable to assume that $\alpha_M$, $\alpha_L$ are proportional to the squared frequencies of the unstable phonon modes. We emphasize that, in Fig.~\ref{fig:dft}(b), we are showing the square root of the absolute value of the negative squared frequency. Therefore, in the experimentally relevant range $-0.4\%\leq (c-c_0)/c_0 \leq 0.4\%$, we conclude that $|\alpha_M|$ and $|\alpha_L|$ have an almost monotonic dependence on $(c-c_0)/c_0$. In particular, $|\alpha_M|$ decreases with increasing $(c-c_0)/c_0$, suggesting that the tensile $c$-axis strain brings the $M$ mode closer to the disordered phase. On the other hand, $|\alpha_L|$ increases with increasing $(c-c_0)/c_0$, indicating that the tensile $c$-axis strain moves the $L$ mode deeper into the ordered phase. Now, experimentally, from Fig. 4(c), we see that $T_{\rm CDW}$ is enhanced by an expanding $c$-axis. One possibility, therefore, is that it is the $L$ mode that is becoming soft at the CDW transition, with the $M$ mode being triggered only due to the trilinear coupling in Eq. (\ref{equ:landau}). The caveat with this scenario is that the $M$ mode seems to generally have a larger (absolute) frequency than the $L$ mode.

Another possibility is that the main effect of the $c$-axis change is not on the phonon frequencies, but on the trilinear coupling $\gamma_{ML}$ of Eq. (\ref{equ:landau}). This scenario seems more likely for several reasons. First, as discussed in Ref. \cite{christensen2021theory}, a relatively large trilinear coupling is needed to obtain a single transition to a $2\times2\times2$ CDW state that breaks sixfold rotational symmetry, as observed experimentally \cite{zhao2021cascade,li2021rotation}. Second, in this scenario $T_{\rm CDW}$ is rather sensitive to $\gamma_{ML}$. Third, because $\gamma_{ML}$ couples order parameters with the same in-plane wave-vector components but different out-of-plane wave-vector components, it seems reasonable to expect that $\gamma_{ML}$ has a pronounced dependence on $c$. As a result, $c$ will be the primary parameter in controlling $T_{\rm CDW}$, as revealed in Fig. 4. Of course, additional DFT calculations are needed to conclusively decide if changes in the parameters $\alpha_{L,M}$ or $\gamma_{ML}$ are responsible for the experimentally observed change in the CDW transition temperature upon uniaxial strain.

The DFT results also provide insight into the mechanism by which $T_C$ increases when the $c$ axis is compressed. The DOS of the high symmetry (i.e. non-CDW) phase of CsV$_3$Sb$_5$ is displayed in Fig.~\ref{fig:dft}(c). By not taking into account the structural symmetry lowering induced by the CDW, we focus only on the direct effect of strain on the DOS. We observe the van Hove singularity (VHS) right below the Fermi level~\cite{kang2021twofold,cho2021emergence,hu2021rich}, which is believed to be important in determining the superconducting instability~\cite{Nandkishore2012,Kiesel2013,Wang2013,wu2021nature}. Since in our calculations no CDW order is considered, if the change in the DOS under strain can account for the change in $T_C$ observed experimentally, this would suggest that the strain-induced enhancement of $T_C$ would happen even in the absence of CDW order, suggesting a weak effect of the CDW phase on the SC properties. However, this does not seem to be the case here. Although a quantitative estimate of $T_C$ is challenging, we see that decreasing the $c$-axis lattice parameter partially suppresses the DOS peak corresponding to the VHS. This is expected to cause a decrease in $T_C$, as opposed to the experimental observation. This analysis corroborates the conclusion drawn from Fig. 4(d) that the changes in $T_C$ under uniaxial strain are dominated by the competition with the CDW state, such that a strain-driven enhancement (suppression) of $T_{\rm CDW}$ results in a suppression (enhancement) of $T_C$. This result, in turn, could be a consequence of the CDW and SC states competing for the same electronic states.

\section{Conclusions}

In summary, we investigated the interplay between CDW and SC in CsV$_3$Sb$_5$ under uniaxial strain applied along the $a$-axis. Comparing our results with recent pressure measurements~\cite{chen2021double,yu2021unusual,zhang2021pressure}, we conclude that both $T_C$ and $T_{\rm CDW}$ are dominated by changes in the $c$-axis lattice parameter, regardless of whether they are promoted by hydrostatic pressure or uniaxial strain. Therefore, this comparison further suggests that the effect of the broken rotational-symmetry induced by the uniaxial strain on the CDW and SC states is weak ~\cite{chen2021double,yu2021unusual}. Moreover, combined with a theoretical analysis, our results not only highlight the importance of the coupling between the two unstable phonon modes on the formation of the CDW, but also indicate that the enhancement of $T_C$ with decreasing $c$ is due to the suppression of the competing CDW instability.


\begin{acknowledgments}
Work at UCLA was supported by the U.S. Department of Energy, Office of Science, Office of Basic Energy Sciences under Award Number DE-SC0021117 for the single crystal growth and measurements under ambient and strained conditions. M.H.C acknowledges support from the Carlsberg foundation. B. M. A. acknowledges support from the Independent Research Fund Denmark grant number 8021-00047B. R.M.F. (phenomenological analysis) was supported by the U.S. Department of Energy, Office of Science, Basic Energy Sciences, Materials Science and Engineering Division, under Award No. DE-SC0020045. T.B. (DFT calculations) was supported by the NSF CAREER grant DMR-2046020.
\end{acknowledgments}


\bibliographystyle{apsrev4-1}
\bibliography{CVS}

\end{document}